%% file: eb2eiff.tex
\documentclass[sigplan,10pt]{acmart}
\usepackage{booktabs}   
\input{prooftree}
\usepackage{bsymb}
\usepackage{subcaption} %% For complex figures with subfigures/subcaptions
\definecolor{myblue4}{RGB}{26,89,142}
\definecolor{kwblue}{RGB}{0,0,128}
\definecolor{eifblue}{RGB}{0,0,255}
\definecolor{eifred}{RGB}{178,34,34}
\definecolor{mygreen}{RGB}{109,223,151}

\newcommand{\eifkw}[1]{\textcolor{kwblue}{{\bf #1}}}
\newcommand{\eif}[1]{\textsf{#1}}
\newcommand{\eiftype}[1]{\eif{#1}}

\newcommand{\eifcomment}[1]{~\textcolor{eifred}{{-}{-}~\textsf{#1}}}

\newcommand{\eiftag}[1]{\textcolor{eifblue}{\text{#1}}}

\newcommand{\ebkeyw}[1]{\textsf{#1}}
\newcommand{\ebtag}[1]{\textsf{\bf{#1}}}

\def \EBN {\textsf{$\delta$}}
\newcommand{\EB}[1]{\EBN\textsf{(}\ensuremath{{#1}}\textsf{)}}

\def \EPN {\textsf{$\xi$}}
\newcommand{\EP}[1]{\EPN\textsf{(}\ensuremath{{#1}}\textsf{)}}

\def \TypeN {\textsf{$\tau$}}
\newcommand{\Type}[1]{\TypeN\textsf{(}\ensuremath{{#1}}\textsf{)}}

\acmConference[2017]{Conf.}
{2017}
{}
\setcopyright{none}             %% For review submission
\begin{document}

%% Title information
%\title{Mapping Event-B machines into Eiffel programming language}
\title{Translating Event-B machines to Eiffel programs}
%% [Short Title] is optional;
                                        %% when present, will be used in
                                        %% header instead of Full Title.
%\titlenote{}             %% \titlenote is optional;
                                        %% can be repeated if necessary;
                                        %% contents suppressed with 'anonymous'
%\subtitle{Taking full advantage of Design-by-Contract}                     %% \subtitle is optional
%\subtitlenote{with subtitle note}       %% \subtitlenote is optional;
                                        %% can be repeated if necessary;
                                        %% contents suppressed with 'anonymous'

%% Author information
%% Contents and number of authors suppressed with 'anonymous'.
%% Each author should be introduced by \author, followed by
%% \authornote (optional), \orcid (optional), \affiliation, and
%% \email.
%% An author may have multiple affiliations and/or emails; repeat the
%% appropriate command.
%% Many elements are not rendered, but should be provided for metadata
%% extraction tools.

%% Author with single affiliation.
\author{Victor Rivera, JooYoung Lee, Manuel Mazzara, Leonard Johard}
%\authornote{with author1 note}          
%% \authornote is optional;
%% can be repeated if necessary
%\orcid{nnnn-nnnn-nnnn-nnnn}
%% \orcid is optional
\affiliation{
  %\department{Software Engineering Lab}              %% \department is recommended
  \institution{Innopolis University}            %% \institution is required
  %\streetaddress{Street1 Address1}
  %\city{Innopolis}
  %\state{State1}
  %\postcode{Post-Code1}
  \country{Russia}
}
\email{v.rivera@innopolis.ru, j.lee@innopolis.ru, m.mazzara@innopolis.ru, l.johard@innopolis.ru}          %% \email is recommended

\begin{abstract}
% Formal modelling languages play a key role in the development of software. Mainly because properties of the systems can be proven correct. However, there is not a clear distinction on how to map the formal model to a programming language. This paper presents a source-to-source mapping between Event-B models to Eiffel programs. Having a model mapped to an Eiffel program is convenient since users can still prove properties of the system (using Design-by-Contract approach, natively supported by Eiffel), while making use of all features of O-O programming.

Formal modelling languages play a key role in the development of software since they enable users to prove correctness of system properties. However, there is still not a clear understanding on how to map a formal model to a specific programming language. In order to propose a solution, this paper presents a source-to-source mapping between Event-B models and Eiffel programs, therefore enabling the proof of correctness of certain system properties via Design-by-Contract (natively supported by Eiffel), while still making use of all features of O-O programming.
\end{abstract}

%% 2012 ACM Computing Classification System (CSS) concepts
%% Generate at 'http://dl.acm.org/ccs/ccs.cfm'.
% <concept_id>10011007.10011006.10011008</concept_id>
% <concept_desc>Software and its engineering~General programming languages</concept_desc>
% <concept_id>10003456.10003457.10003521.10003525</concept_id>
% <concept_desc>Social and professional topics~History of programming languages</concept_desc>
% <concept_significance>300</concept_significance>

%\ccsdesc[500]{Software and its engineering~General programming languages}
%\ccsdesc[300]{Social and professional topics~History of programming languages}
%% End of generated code

%% Keywords
%% comma separated list
%% \keywords is optional

%% \maketitle
%% Note: \maketitle command must come after title commands, author
%% commands, abstract environment, Computing Classification System
%% environment and commands, and keywords command.
\maketitle

\section{Introduction}
The importance of developing correct software systems has been increased in the past few years. Final users of systems trust systems and are not aware of the consequences of malfunctioning. Hence, the burden is on developers, engineers and researchers that have to pay close attention to the development of flawless systems. There are different approaches to tackle the problem, e.g. top-down and bottom-up approaches: using a top-down approach, one could think to start developing the system from a very abstract view point towards more concrete ones; in a bottom-up approach, on the other hand, one might think to start from a more concrete state of the system to then add more functionality to it. The key point on both approaches is to always prove that properties of the systems hold.

Event-B is a formal modelling language for reactive systems,
introduced by Abrial \cite{Abrial:EB:Book:10}, which allows the modelling of complete systems. It follows the top-down approach by means of refinements. One can create an abstraction of the system and express its properties. Prove that the system indeed meets the properties to then create a refinement of the system: same system with more details. It has been applied with success in both research and industrial projects, and in integrated EU projects aiming at putting together the two dimensions, for example in the automotive sector \cite{GmehlichGLIJM13}.

On the other side of the spectrum, following a bottom-up approach, one can work with Eiffel programming language \cite{Meyer:1992}. In Eiffel, one can create classes that implement any system. The behaviour of such classes is specified in Eiffel using contracts: pre- and post-conditions and class invariants. These mechanisms are natively supported by the language. Having contracts, one can then verify that the implementation is indeed the intended. After the implementation of the class, one can give more speciality or generalization by using inheritance. This paper gives a series of rules to generate Eiffel programs from Event-B model, bridging both top-down and bottom-up approaches. Rules take into account system specifications of the Event-B model and generate either Eiffel code or contracts. Thus, users will end up with an implementation of the system while they can prove it correct.

%\section{Related Work}
Several translations have been achieved that go in the same direction as the work presented on this paper. In \cite{Mery:2011}, M\`ery and Singh present the \texttt{EB2ALL} tool-set that includes a translation from Event-B models to C, C++ and Java. Unlike this translation, \texttt{EB2ALL} provides support for a small part of Event-B's syntax, and users are required to write a final Event-B implementation refinement in the syntax supported by the tool. The Code Generation tool \cite{soton2010} generates concurrent Java and Ada programs for a tasking extension of Event-B. Unlike  these tools, the work presented here does not require user's intervention, while it works on the proper syntax of the Event-B model. In addition, these tools do not take full advantage of the elements present in the source language, e.g. invariants. The work presented in this paper, in addition to an implementation, generates contracts from the source language, making use of the Design-by-Contract approach. In \cite{Rivera:2017,Catano:2016}, authors present a translation from Event-B to Java, annotating the code with JML (Java Modelling Language) specifications, and \cite{rivera:tok:16} shows its application. The main difference with the work presented here is the target language. We are translating to Eiffel which natively supports Design-by-Contract. In addition, Eiffel comes with different tools to statically prove Eiffel code (e.g. Autoproof \cite{TFNP-TACAS15}) that fully supports the language. Another difference is the translation of carrier sets. EventB2Java translates them as set of integers 

\section{Preliminaries}
\subsection{Event-B}
Event-B is a formal modelling language for reactive systems,
introduced by Abrial \cite{Abrial:EB:Book:10}, which allows the modelling of complete systems. Figure \ref{general} shows the general view of an Event-B machine and context. Event-B models are composed of contexts and machines. Contexts define constants (written after \ebkeyw{constant} in context $C$), uninterpreted sets (written after \ebkeyw{set} in context $C$) and their properties (written after \ebkeyw{axioms} in context $C$). Machines define variables (written after \ebkeyw{variables} in machine $M$) and their properties (expressed as invariants after \ebkeyw{invariant} in machine $M$), and state transitions expressed as events (written between \ebkeyw{events} and the last \ebkeyw{end}). The initialisation event gives initial values to variables. 
\begin{figure}
{
\[ 
    \begin{array}{ll}
    \begin{array}{l}
	\ebkeyw{machine }M \ebkeyw{ sees } C \\
    \ebkeyw{variables}\;v \\
    \ebkeyw{invariants}\;label\_inv:\;I(s,c,v) \\
    \ebkeyw{events}\\
    \hspace{.3cm}\ebkeyw{event}\;initialisation\\
    \hspace{.5cm}\ebkeyw{then}\:A(s,c,v) \:\ebkeyw{end} \\
    \hspace{.3cm} \ebkeyw{event}\;evt \\
    \hspace{.5cm}\ebkeyw{any}\;x\\  
    \hspace{.3cm}\ebkeyw{where}\\  
    \hspace{.5cm}label\_guard:\;G(s, c, v, x)\\  
    \hspace{.3cm}\ebkeyw{then}\\  
    \hspace{.5cm}label\_action:\;A(s, c, v, x)\\  
    \hspace{.3cm}\ebkeyw{end}\\
    \ebkeyw{end}
    \end{array}
    &
    \begin{array}{l}
	\ebkeyw{Context }\;C \\
    \ebkeyw{constant}\;c \\
    \ebkeyw{set}\;S \\
    \ebkeyw{axioms}\;X (s, c) \\
    \ebkeyw{end}
    \end{array}
    \end{array}
\]
}
\caption{General view of an Event-B machine and its context.}
\label{general}
\end{figure}

An event is composed of guards and actions. The guard (written between keywords \ebkeyw{where} and \ebkeyw{then}) represents
conditions that must hold for the event to be
triggered. The action (written between keywords \ebkeyw{then} and \ebkeyw{end}) gives new values to variables

In Event-B, systems are modelled via a sequence
of refinements. First, an abstract machine is developed and
verified to satisfy whatever correctness and safety properties
are desired. Refinement machines are used to add more detail
to the abstract machine until the model is sufficiently concrete
for hand or automated translation to code. Refinement proof
obligations are discharged to ensure that each refinement is a
faithful model of the previous machine, so that all machines
satisfy the correctness properties of the original.

\subsection{Eiffel}
Eiffel is an Object-Oriented programming language that natively supports the Design-by-Contract methodology. The behaviour of classes is specified by equipping them with contracts. Each routine of the class contains a pre- and post-condition: a client of a routine needs to guarantee the pre-condition on routine call. In return, the post-condition of the procedure, on routine exit, holds. The class is also equipped with class invariants. Invariants maintain the consistency of objects. Contracts in Eiffel follow a similar semantics of Hoare Triples.

Figure \ref{fig:class} depicts an Eiffel class that implements part of a Bank Account. The name of the class is \eif{ACCOUNT} and it appears right after the keyword \eifkw{class}. In Eiffel, implementers need to list creation procedures after the keyword \eifkw{create}. In Figure \ref{fig:class}, \eif{make} is a procedure of the class that can be used as a creation procedure. Class \eif{ACCOUNT} structures its procedures in \texttt{Initialisation}, \texttt{Access} and \texttt{Element change}, by using the keyword \eifkw{feature}. This structure can be use for information hiding (not discussed here). \eif{balance} is a class attribute that contains the actual balance of the account. It is defined as an integer. Procedures in Eiffel are defined by given them a name (e.g. \eif{withdraw}) and its respective arguments. It is followed by a head comment (which is optional). Procedures are equipped with pre- and post-conditions predicates. In Eiffel, a predicate is composed of a tag (optional) and a boolean expression. For instance, the pre-condition for \eif{withdraw} (after the key work \eifkw{require}) imposes the restriction on callers to provide and argument that is greater than or equal zero and less than or equal the balance of the account (\eiftag{amount\_not\_negative} and \eiftag{amount\_available} are tags, identifiers, and are optionals). If the pre-condition of the procedure is met, the post-condition  (after the key work \eifkw{ensure}) holds on procedure exit. In a post-condition, the aid \eifkw{old} refers to the value of an expression on procedure entry. The actions of the procedure are listed in between the key words \eifkw{do} and \eifkw{ensure}. The only action of \eif{withdraw} procedure is to increase the value of \eif{balance} by \eif{amount}. Finally, The invariant is restricting the possible values for variables. 
\begin{figure}[h]
{
\[
\begin{array}{l}
\eifkw{class }\; \eif{ACCOUNT}\;\eifkw{create }\;\eif{make}\\
\eifkw{feature} \eifcomment{ Initalisation}\\
\hspace*{.3cm}\eif{make}\\
\hspace*{.9cm}\eifcomment{Initialise an empty account.}\\
\hspace*{.6cm}\eifkw{do}\\
\hspace*{.9cm}\eif{balance := 0}\\
%\hspace*{.9cm}\eif{credit\_limit := 0}\\
\hspace*{.6cm}\eifkw{ensure}\\
\hspace*{.9cm}\eiftag{balance\_set: } \eif{balance = 0}\\
%\hspace*{.9cm}\eiftag{credit\_limit\_set:}\eif{ credit\_limit = 0}\\
\hspace*{.6cm}\eifkw{end}\\
\eifkw{feature} \eifcomment{ Access}\\
\hspace*{.3cm}\eif{balance: INTEGER}\\
\hspace*{.6cm}\eifcomment{Balance of this account.}\\
%\hspace*{.3cm}\eif{credit\_limit: INTEGER}\\
%\hspace*{.6cm}\eifcomment{Credit limit of this account.}\\
%\hspace*{.3cm}\eif{available\_amount: INTEGER}\\
%\hspace*{.9cm}\eifcomment{Amount available on this account.}\\
%\hspace*{.3cm}\eifkw{do}\\
%\hspace*{.9cm}\eifkw{Result}\eif{ := balance + credit\_limit}\\
%\hspace*{.3cm}\eifkw{end}\\
\eifkw{feature} \eifcomment{ Element change}\\
\hspace*{.3cm}\eif{withdraw (amount: INTEGER)}\\
\hspace*{.9cm}\eifcomment{Withdraw `amount' from this account.}\\
\hspace*{.6cm}\eifkw{require}\\
\hspace*{.9cm}\eiftag{amount\_not\_negative: }\eif{amount $>=$ 0}\\
\hspace*{.9cm}\eiftag{amount\_available: }\eif{amount $<=$ balance}\\
\hspace*{.6cm}\eifkw{do}\\
\hspace*{.9cm}\eif{balance := balance - amount}\\
\hspace*{.6cm}\eifkw{ensure}\\
\hspace*{.9cm}\eiftag{balance\_set: }\eif{balance = \eifkw{old} balance - amount}\\
\hspace*{.6cm}\eifkw{end}\\
\eifkw{invariant}\\
\hspace*{.3cm}\eiftag{balance\_not\_negative: }\eif{balance $>=$ 0}\\
%\hspace*{.3cm}\eiftag{balance\_not\_below\_credit: }\eif{balance $>=$ -credit\_limit}\\
\eifkw{end}
\end{array}
\]
} 
\caption{Eiffel class}
\label{fig:class}
\end{figure} 

\section{Translation}
\label{trans}
The translation is done by the aid \EBN\ $:$ \texttt{Event-B} $\tfun$ \texttt{Eiffel}. \EBN\ takes an Event-B model and produces Eiffel classes. It is defined as a total function (i.e. $\tfun$) since any Event-B model can be translated to Eiffel. It uses two helpers:  \EPN\ translates Event-B Expressions or Predicates to Eiffel, and \TypeN\ translates the type of Event-B variable to the corresponding type in Eiffel.

\subsection{Translating Event-B machines}
Rule \texttt{machine} is a high level translation. It takes an Event-B machine \textsf{M} and produces an  Eiffel class \eif{M}.
{
\[ 
\begin{tabular}{cc}
  \begin{prooftree}
    \begin{array}{lll}
		\Type{v} = \texttt{Type}&
        \EP{I(s, c, v)} = \texttt{Inv} &
        \EB{\ebkeyw{events}\: e} = \texttt{E} \\
      \multicolumn{3}{l}{\EB{\ebkeyw{event}\ initialisation\:\ebkeyw{then}\:A(s,c,v) \:\ebkeyw{end}} = \texttt{Init}} \\
    \end{array}
    \using \texttt{(machine)}
    \justifies
    \begin{array}{l}
      \EBN(\ebkeyw{machine }M \ebkeyw{ sees } C \\
      \hspace{1cm} \ebkeyw{variables}\;v \\
      \hspace{1cm} \ebkeyw{invariants}\;label\_inv:\;I(s,c,v) \\
      \hspace{1cm} \ebkeyw{event}\
      initialisation\:\ebkeyw{then}\:A(s,c,v) \:\ebkeyw{end} \\
      \hspace{1cm} \ebkeyw{events}\;e \\
      \hspace{.5cm}\ebkeyw{end}) =\\
      \eifkw{class}\; \eif{M}\;\eifkw{create}\; \eif{initialisation}\\
      \eifkw{feature}\; \eifcomment{Initialisation}\\
      \hspace{.5cm} \texttt{Init}\\
      \eifkw{feature}\; \eifcomment{Events}\\
      \hspace{.5cm} \texttt{E}\\
      \eifkw{feature}\; \eifcomment{Access}\\
      \hspace{.5cm} \eif{ctx}\;:\; \eiftype{CONSTANTS}\\
      \hspace{.5cm}\eif{v}\;:\;\texttt{Type}\\ 
      \eifkw{invariant}\\
      \hspace{.5cm}\eif{label\_inv:}\;\texttt{Inv}\\
      \eifkw{end}
    \end{array}
  \end{prooftree}
\end{tabular} 
\]
}
Variables are translated as class attributes in class \eif{M}. Event-B invariants are translated to Eiffel invariants. Both, Event-B and Eiffel, have similar semantics for invariants. Rule \texttt{context} generate an Eiffel class \eif{CONSTANT} that contains the translation of Event-B constants and carrier sets defined by the user. Axioms, which restrict the possible values for constants are translated to invariants of this class. Constants in Event-B are entities that cannot change their values. They are naturally translated to Eiffel as \eifkw{once} variables.
{
\[ 
\begin{tabular}{cc}
  \begin{prooftree}
    \begin{array}{ll}
      \EB{\ebkeyw{axioms}\;X(s, c)} = \texttt{X}\\
      \Type{c} = \texttt{Type}
    \end{array}
    \using \texttt{(context)}
    \justifies
    \begin{array}{l}
      \EBN(\ebkeyw{Context }\;C \\
      \hspace{1cm} \ebkeyw{constant}\;c \\
      \hspace{1cm} \ebkeyw{set}\;S \\
      \hspace{1cm} \ebkeyw{Axioms}\;X (s, c) \\
      \hspace{.5cm}\ebkeyw{end}) =\\
      \eifkw{class}\; \eif{CONSTANTS}\\
      %	\eifkw{create}\; \eif{make}\\ \\
      \eifkw{feature} \eifcomment{Constants}\\
      \hspace{.5cm} \eif{c}:\; \texttt{Type}\\
      \hspace*{1.5cm} \eifcomment{`c' comment}\\
      \hspace{1cm} \eifkw{once}\\
      \hspace{1.5cm} \eifkw{create}\; \texttt{Type} \;\eifkw{Result}\\
      \hspace{1cm} \eifkw{end}\\
      \eifkw{invariant}\\
      \hspace{.5cm}\texttt{X}\\
      \eifkw{end}
    \end{array}
  \end{prooftree}
\end{tabular} 
\]
}
Carrier sets represent a new type defined by the user. Each carrier set is  translated as an afresh Eiffel class so users are able to use them as types. Rule \texttt{cset} shows the translation. Parts of the class are omitted due to space. Class \eif{EBSET [T]} gives an implementation to sets of type \eif{T}. Class \eif{S} inherits \eif{EBSET [T]} due to the nature of carrier sets in Event-B.

{
\[ 
\begin{tabular}{cc}
  \begin{prooftree}
    \begin{array}{ll}
      \Type{s} = \texttt{Type}
    \end{array}
    \using \textsf{(cset)}
    \justifies
    \begin{array}{l}
      \EBN(\ebkeyw{Context }\;C \\
      \hspace{1cm} \ebkeyw{constant}\;c \\
      \hspace{1cm} \ebkeyw{set}\;S \\
      \hspace{1cm} \ebkeyw{Axioms}\;X (s, c) \\
      \hspace{.5cm}\ebkeyw{end}) = \\
      \eifkw{class}\; S\\
    \eifkw{inherit}\\
    \hspace{.5cm}\eiftype{EBSET}\; [\texttt{Type}]\\
    	\ldots \\ 
    \eifkw{end}
    \end{array}
  \end{prooftree}
\end{tabular} 
\]
}
Rule \texttt{event} produces an Eiffel feature given an Event-B $event$. Parameters of the event are translated as arguments of the respective feature in Eiffel with its respective type. In Event-B, an event might be executed only if the guard is true. In Eiffel, the guard is translated as the precondition of the feature. Hence, the client is now in charge of meeting the specification before calling the feature. The semantics of the execution is handle now by the client who wants to execute the feature rather than the system deciding. The actual execution of the actions still preserve its semantics: execution of the actions is only possible if the guard is true. In Eiffel, for a client to execute a feature he needs to meet the guard otherwise a runtime exception will be raised: Contract violation. 

Event-B event actions are translated directly to Eiffel statements. In Event-B, the before-after predicate contains primed and unprimed variables representing the before and after value of the variables. We translated the primed variable with the Eiffel key word \eifkw{old}. Representing old value of the variable. For simplicity. the rule only takes into account a single parameter, a single guard and a single action. However, this can be easily extended.
{
\[ 
\begin{tabular}{cc}
  \begin{prooftree}
    \begin{array}{ll}
      \EP{G(s, c, v, x)} = \texttt{G}&
                                       \EP{A(s, c, v, x)} = \texttt{A}\\
      \Type{x} = \texttt{Type}
    \end{array}
    \using \texttt{(event)}
    \justifies
    \begin{array}{l}
      \EBN(\ebkeyw{event}\ evt\;\ebkeyw{any}\;\ebkeyw{x}\\  
      \hspace{1cm}\ebkeyw{where}\;label\_guard:\;G(s, c, v, x)\\  
      \hspace{1cm}\ebkeyw{then}\;label\_action:\;A(s, c, v, x)\\  
      \hspace{1cm}\ebkeyw{end}) =\\ 
      \hspace{.5cm}\eif{evt} (\eif{x}: \texttt{Type})\\ 
      \hspace*{1.5cm}\eifcomment{'evt' comment } \\ 
      \hspace{1cm}\eifkw{require} \\
      \hspace{1.5cm}\eif{label\_guard:}\;\texttt{G} \\ 
      \hspace{1cm}\eifkw{do} \\ 
      \hspace{1.5cm}\eif{v.assigns(}\texttt{A}\eif{)} \\ 
      \hspace{1cm}\eifkw{ensure} \\
      \hspace{1.5cm}\eif{label\_action:}\;\eif{v.equals(}\eifkw{old}\;\texttt{A}\eif{)}\\ 
      \hspace{1cm}\eifkw{end} \\  \\ 
    \end{array}
  \end{prooftree}
  % TODO
  % some parenthesis are no with eif
  % to put the label in the EB code
  
\end{tabular} 
\]
}

Rule \texttt{init} below shows the translation of Event-B event $initialisation$ to a creation procedure in Eiffel. The creation procedure initialises the object containing the constants definition. It also assigns initial values to variables taken from the initialisation in the $initialisation$ event. In Eiffel, creation procedures are listed under the keyword \eifkw{create}, as shown in rule \texttt{machine}. The \eifkw{ensure} clause shows the translation of the before-after predicate of the assignment in Event-B.
{
\[ 
\begin{tabular}{cc}
  \begin{prooftree}
    \begin{array}{ll}
      \EP{A(s, c, v)} = \texttt{A}\\
    \end{array}
    \using \texttt{(init)}
    \justifies
    \begin{array}{l}
      \EBN(\ebkeyw{event}\ initialisation\\
      \hspace{1cm}\ebkeyw{then}\\
      \hspace{1.5cm}label:\;A(s,c,v) \\
      \hspace{.5cm}\ebkeyw{end}) =\\ \\
      \hspace{.5cm} \eif{initialisation}\\
      \hspace*{1.5cm} \eifcomment{evt comment}\\
      \hspace{1cm} \eifkw{do}\\
      \hspace{1.5cm} \eifkw{create}\; \eif{ctx}\\
      \hspace{1.5cm} \eif{v.assigns}(\texttt{A})\\
      \hspace{1cm} \eifkw{ensure}\\
      \hspace{1.5cm} \eif{label: v.is\_equal}(\eifkw{old}\;\texttt{A})\\
      \hspace{1cm} \eifkw{end}\\
    \end{array}
  \end{prooftree}
  
\end{tabular} 
\]
}

\subsection{Hand translation}
In this Section, we apply (manually) the translation rules to the Event-B model in Figure \ref{eb-model}. The Event-B model is a well known model created by Abrial in \cite{Abrial:EB:Book:10}. It models a system for controlling cars in an island and on a bridge. The model depicted in Figure \ref{eb-model} only shows the most abstract model of the system.

\begin{figure}
{
\[
\begin{array}{ll}
 \begin{array}{l}
	\ebkeyw{machine}\; m0\; \ebkeyw{sees}\; c0\\
    \ebkeyw{variables}\; n\\
    \ebkeyw{invariants}\\
  	\hspace{.5cm}\ebtag{inv1: }  n \in \nat\\
	\hspace{.5cm}\ebtag{inv2: } n\le d\\
  	%\hspace{.5cm}\ebtag{thm1: } n>0 \vee n<d\\
	\ebkeyw{events}\\
  	\hspace{.5cm}\ebkeyw{event } INITIALISATION\\
    \hspace{1cm}\ebkeyw{then}\\
    \hspace{1.5cm}\ebtag{act1 } n := 0\\
  	\hspace{.5cm}\ebkeyw{end}\\

  	\hspace{.5cm}\ebkeyw{event }ML\_out\\
    \hspace{1cm}\ebkeyw{where}\\
    \hspace{1.5cm}\ebtag{grd1 } n<d\\
    \hspace{1cm}\ebkeyw{then}\\
    \hspace{1.5cm}\ebtag{act1 } n := n+1\\
  	\hspace{.5cm}\ebkeyw{end}\\

	\hspace{.5cm}\ebkeyw{event } ML\_in\\
	\hspace{1cm}\ebkeyw{where}\\
    \hspace{1.5cm}\ebtag{grd1 }n>0\\
    \hspace{1cm}\ebkeyw{then}\\
    \hspace{1.5cm}\ebtag{act1 }n : =n-1\\
	\hspace{.5cm}\ebkeyw{end}\\
	\ebkeyw{end}\\
    \end{array}
    &
    \begin{array}{l}
	\ebkeyw{context }c0\\
	\ebkeyw{constants }d \\
    \ebkeyw{axioms}\\
	\hspace{.5cm}\ebtag{axm1 }d\in\nat\\
	\hspace{.5cm}\ebtag{axm2 }d>0\\
	\ebkeyw{end}
    \end{array}
    \end{array}
\]
}
\caption{Controlling cars on a bridge: Event-B machine and its context.}
\label{eb-model}
\end{figure}

Machine $m0$ sees context $c0$. $c0$ defines a constant $d$ as a natural number greater than 0. This constant models the maximum number of cars that can be on the island and bridge. Machine $m0$ also defines a variable $n$ as a natural number (predicate \ebtag{inv1}). Variable $n$ is the actual number of cars in the island and on the bridge. Predicate \ebtag{inv2} imposes the restriction on the number of cars, it must not be over $d$. Event $initialisation$ gives an initial value to $n$: no cars in the island or on the bridge. Event $ML\_out$ models the transition for a car in the mainland to enter the island. The restriction is that the number of cars already in the island is strictly less than $d$: there is room for at least another car. Its action is to increase the number of cars in the island by one. Event $ML\_in$ models the transition for a car in the island to enter the mainland. The only restriction is that there is at least one car in the island. Its action is to decrease the number of cars in the island. All these restrictions are ensured by the proof obligations.

Figure \ref{translation} is the mapping to Eiffel programming language by applying the rules in Section \ref{trans}. 

\begin{figure}
{\scriptsize
\[
\begin{array}{ll}
 \begin{array}{l}
\eifkw{class }\; \eif{m0}\;\eifkw{create }\;\eif{INITIALISATION}\\
\eifkw{feature} \eifcomment{ Initalisation}\\
\hspace*{.3cm}\eif{initialisation}\\
\hspace*{.6cm}\eifkw{do}\\
\hspace*{.9cm}\eifkw{create}\;\eif{ctx}\\
\hspace*{.9cm}\eif{n := 0}\\
\hspace*{.6cm}\eifkw{ensure}\\
\hspace*{.9cm}\eiftag{act1: } \eif{n = 0}\\
\hspace*{.6cm}\eifkw{end}\\
\eifkw{feature} \eifcomment{ Events}\\
\hspace*{.3cm}\eif{ml\_out}\\
\hspace*{.6cm}\eifkw{require}\\
\hspace*{.9cm}\eiftag{grd1: }\eif{n $<$ d}\\
\hspace*{.6cm}\eifkw{do}\\
\hspace*{.9cm}\eif{n := n + 1}\\
\hspace*{.6cm}\eifkw{ensure}\\
\hspace*{.9cm}\eiftag{act1: }\eif{n = \eifkw{old} n + 1}\\
\hspace*{.6cm}\eifkw{end}
    \end{array}
    &
    \begin{array}{l}
\hspace*{.3cm}\eif{ml\_in}\\
\hspace*{.6cm}\eifkw{require}\\
\hspace*{.9cm}\eiftag{grd1: }\eif{n $>$ 0}\\
\hspace*{.6cm}\eifkw{do}\\
\hspace*{.9cm}\eif{n := n - 1}\\
\hspace*{.6cm}\eifkw{ensure}\\
\hspace*{.9cm}\eiftag{act1: }\eif{n = \eifkw{old} n - 1}\\
\hspace*{.6cm}\eifkw{end}\\
\eifkw{feature} \eifcomment{ Access}\\
\hspace*{.3cm}\eifkw{ctx: CONSTANTS}\\
\hspace*{.3cm}\eifkw{n: INTEGER}\\
\eifkw{invariant}\\
\hspace*{.3cm}\eiftag{inv1: }\eif{n $>=$ 0}\\
\hspace*{.3cm}\eiftag{inv2: }\eif{n $<=$ d}\\
\eifkw{end}
\end{array}
    \end{array}
\]
}
\caption{Excerpt of the Eiffel translation from the Event-B model depicted in Figure \ref{eb-model}.}
\label{translation}
\end{figure}

\section{Conclusion}
We presented a series of rules to transform an Event-B model to an Eiffel program. The translation takes full advantage of all elements in the source by translating them as contracts in the target language. Thus, no information on the behaviour of the system is lost. These rules shows a methodology for software construction that makes use of two different approaches. 

We plan on implementing these rules as an Event-B plug-in. We also plan of taking full advantage of the Proof Obligations generated by Event-B: translated them into a specification driven class so to help Eiffel provers in the process of proving the correctness of classes after any modification (extension) done by the implementer.

%non-deterministic: deferred

%% Acknowledgments
%\begin{acks}
%% acks environment is optional
%% contents suppressed with 'anonymous'
%% Commands \grantsponsor{<sponsorID>}{<name>}{<url>} and
%% \grantnum[<url>]{<sponsorID>}{<number>} should be used to
%% acknowledge financial support and will be used by metadata
%% extraction tools.
%  This material is based upon work supported by the
%  \grantsponsor{GS100000001}{National Science
 %   Foundation}{http://dx.doi.org/10.13039/100000001} under Grant
%  No.~\grantnum{GS100000001}{nnnnnnn} and Grant
%  No.~\grantnum{GS100000001}{mmmmmmm}.  Any opinions, findings, and
%  conclusions or recommendations expressed in this material are those
%  of the author and do not necessarily reflect the views of the
%  National Science Foundation.
%\end{acks}

%% Bibliography
%% Bibliography style
\bibliographystyle{ACM-Reference-Format}
\bibliography{eb2eiff}

%% Appendix
%\appendix
%\section{Appendix}

%Text of appendix \ldots

\end{document}

%% file: prooftree.tex
\newdimen\proofrulebreadth \proofrulebreadth=.05em
\newdimen\proofdotseparation \proofdotseparation=1.25ex
\newdimen\proofrulebaseline \proofrulebaseline=2ex
\newcount\proofdotnumber \proofdotnumber=3
\let\then\relax
\def\hfi{\hskip0pt plus.0001fil}
\mathchardef\squigto="3A3B
\newif\ifinsideprooftree\insideprooftreefalse
\newif\ifonleftofproofrule\onleftofproofrulefalse
\newif\ifproofdots\proofdotsfalse
\newif\ifdoubleproof\doubleprooffalse
\let\wereinproofbit\relax
\newdimen\shortenproofleft
\newdimen\shortenproofright
\newdimen\proofbelowshift
\newbox\proofabove
\newbox\proofbelow
\newbox\proofrulename
\def\shiftproofbelow{\let\next\relax\afterassignment\setshiftproofbelow\dimen0 }
\def\shiftproofbelowneg{\def\next{\multiply\dimen0 by-1 }%
\afterassignment\setshiftproofbelow\dimen0 }
\def\setshiftproofbelow{\next\proofbelowshift=\dimen0 }
\def\setproofrulebreadth{\proofrulebreadth}

\def\prooftree{% NESTED ZERO (\ifonleftofproofrule)
\ifnum  \lastpenalty=1
\then   \unpenalty
\else   \onleftofproofrulefalse
\fi
\ifonleftofproofrule
\else   \ifinsideprooftree
        \then   \hskip.5em plus1fil
        \fi
\fi
\bgroup% NESTED ONE (\proofbelow, \proofrulename, \proofabove,
\setbox\proofbelow=\hbox{}\setbox\proofrulename=\hbox{}%
\let\justifies\proofover\let\leadsto\proofoverdots\let\Justifies\proofoverdbl
\let\using\proofusing\let\[\prooftree
\ifinsideprooftree\let\]\endprooftree\fi
\proofdotsfalse\doubleprooffalse
\let\thickness\setproofrulebreadth
\let\shiftright\shiftproofbelow \let\shift\shiftproofbelow
\let\shiftleft\shiftproofbelowneg
\let\ifwasinsideprooftree\ifinsideprooftree
\insideprooftreetrue
\setbox\proofabove=\hbox\bgroup$\displaystyle % NESTED TWO
\let\wereinproofbit\prooftree
\shortenproofleft=0pt \shortenproofright=0pt \proofbelowshift=0pt
\onleftofproofruletrue\penalty1
}

\def\eproofbit{% NESTED TWO
\ifx    \wereinproofbit\prooftree
\then   \ifcase \lastpenalty
        \then   \shortenproofright=0pt  % 0: some other object, no indentation
        \or     \unpenalty\hfil         % 1: empty hypotheses, just glue
        \or     \unpenalty\unskip       % 2: just had a tree, remove glue
        \else   \shortenproofright=0pt  % eh?
        \fi
\fi
\global\dimen0=\shortenproofleft
\global\dimen1=\shortenproofright
\global\dimen2=\proofrulebreadth
\global\dimen3=\proofbelowshift
\global\dimen4=\proofdotseparation
\global\count255=\proofdotnumber
$\egroup  % NESTED ONE
\shortenproofleft=\dimen0
\shortenproofright=\dimen1
\proofrulebreadth=\dimen2
\proofbelowshift=\dimen3
\proofdotseparation=\dimen4
\proofdotnumber=\count255
}

\def\proofover{% NESTED TWO
\eproofbit % NESTED ONE
\setbox\proofbelow=\hbox\bgroup % NESTED TWO
\let\wereinproofbit\proofover
$\displaystyle
}%
\def\proofoverdbl{% NESTED TWO
\eproofbit % NESTED ONE
\doubleprooftrue
\setbox\proofbelow=\hbox\bgroup % NESTED TWO
\let\wereinproofbit\proofoverdbl
$\displaystyle
}%
\def\proofoverdots{% NESTED TWO
\eproofbit % NESTED ONE
\proofdotstrue
\setbox\proofbelow=\hbox\bgroup % NESTED TWO
\let\wereinproofbit\proofoverdots
$\displaystyle
}%
\def\proofusing{% NESTED TWO
\eproofbit % NESTED ONE
\setbox\proofrulename=\hbox\bgroup % NESTED TWO
\let\wereinproofbit\proofusing
\kern0.3em$
}

\def\endprooftree{% NESTED TWO
\eproofbit % NESTED ONE
  \dimen5 =0pt% spread of hypotheses
\dimen0=\wd\proofabove \advance\dimen0-\shortenproofleft
\advance\dimen0-\shortenproofright
\dimen1=.5\dimen0 \advance\dimen1-.5\wd\proofbelow
\dimen4=\dimen1
\advance\dimen1\proofbelowshift \advance\dimen4-\proofbelowshift
\ifdim  \dimen1<0pt
\then   \advance\shortenproofleft\dimen1
        \advance\dimen0-\dimen1
        \dimen1=0pt
        \ifdim  \shortenproofleft<0pt
        \then   \setbox\proofabove=\hbox{%
                        \kern-\shortenproofleft\unhbox\proofabove}%
                \shortenproofleft=0pt
        \fi
\fi
\ifdim  \dimen4<0pt
\then   \advance\shortenproofright\dimen4
        \advance\dimen0-\dimen4
        \dimen4=0pt
\fi
\ifdim  \shortenproofright<\wd\proofrulename
\then   \shortenproofright=\wd\proofrulename
\fi
\dimen2=\shortenproofleft \advance\dimen2 by\dimen1
\dimen3=\shortenproofright\advance\dimen3 by\dimen4
\ifproofdots
\then
        \dimen6=\shortenproofleft \advance\dimen6 .5\dimen0
        \setbox1=\vbox to\proofdotseparation{\vss\hbox{$\cdot$}\vss}%
        \setbox0=\hbox{%
                \advance\dimen6-.5\wd1
                \kern\dimen6
                $\vcenter to\proofdotnumber\proofdotseparation
                        {\leaders\box1\vfill}$%
                \unhbox\proofrulename}%
\else   \dimen6=\fontdimen22\the\textfont2 % height of maths axis
        \dimen7=\dimen6
        \advance\dimen6by.5\proofrulebreadth
        \advance\dimen7by-.5\proofrulebreadth
        \setbox0=\hbox{%
                \kern\shortenproofleft
                \ifdoubleproof
                \then   \hbox to\dimen0{%
                        $\mathsurround0pt\mathord=\mkern-6mu%
                        \cleaders\hbox{$\mkern-2mu=\mkern-2mu$}\hfill
                        \mkern-6mu\mathord=$}%
                \else   \vrule height\dimen6 depth-\dimen7 width\dimen0
                \fi
                \unhbox\proofrulename}%
        \ht0=\dimen6 \dp0=-\dimen7
\fi
\let\doll\relax
\ifwasinsideprooftree
\then   \let\VBOX\vbox
\else   \ifmmode\else$\let\doll=$\fi
        \let\VBOX\vcenter
\fi
\VBOX   {\baselineskip\proofrulebaseline \lineskip.2ex
        \expandafter\lineskiplimit\ifproofdots0ex\else-0.6ex\fi
        \hbox   spread\dimen5   {\hfi\unhbox\proofabove\hfi}%
        \hbox{\box0}%
        \hbox   {\kern\dimen2 \box\proofbelow}}\doll%
\global\dimen2=\dimen2
\global\dimen3=\dimen3
\egroup % NESTED ZERO
\ifonleftofproofrule
\then   \shortenproofleft=\dimen2
\fi
\shortenproofright=\dimen3
\onleftofproofrulefalse
\ifinsideprooftree
\then   \hskip.5em plus 1fil \penalty2
\fi
}